\documentclass[aps,twocolumn,amsmath,amssymb,reprint,numbers,superscriptaddress,noeprint,longbibliography]{revtex4-1}

\usepackage{soul}
\usepackage{geometry}
\usepackage{graphicx}
\usepackage{booktabs}
\usepackage{array}
\usepackage{textcase}
\usepackage{relsize}
\usepackage{color}
\usepackage{xcolor}
\usepackage{eqnarray}
\usepackage{caption}
\usepackage{textcomp}
\usepackage{setspace}
\usepackage{bm}
\usepackage[normalem]{ulem}

\usepackage{subfig}
\usepackage{tikz}
\usetikzlibrary{positioning}
\usepackage{hyperref}
\hypersetup{
     colorlinks   = true,
     linkcolor    = blue,
     citecolor    = blue,
     urlcolor = blue
}

\definecolor{cola}{rgb}{0.7,0.1,0.1}
\definecolor{colb}{rgb}{0.9,0.4,0}
\definecolor{colc}{rgb}{0.3,0.7,0}
\definecolor{cold}{rgb}{0,0.35,0.75}
\definecolor{cole}{rgb}{0.63, 0.13, 0.94}
\definecolor{colf}{rgb}{0.63, 0.13, 0.13}
\definecolor{colg}{rgb}{0, 0.6, .6}

\newcommand{\addMisha}[1]{{\color{colf} #1}}

\begin{document}

\title[]{Microscopic model of stacking-fault potential and exciton wave function in GaAs}

\author{Mikhail V. Durnev}
\affiliation{Ioffe Institute, 194021 St.-Petersburg, Russia}
\author{Mikhail M. Glazov}
\affiliation{Ioffe Institute, 194021 St.-Petersburg, Russia}
\author{Xiayu Linpeng}
\affiliation{Department of Physics, University of Washington, Seattle, Washington 98195, USA}
\author{Maria L. K. Viitaniemi}
\affiliation{Department of Physics, University of Washington, Seattle, Washington 98195, USA}
\author{Bethany Matthews}
\author{Steven R. Spurgeon}
\affiliation{Energy and Environment Directorate, Pacific Northwest National Laboratory, Richland, Washington 99352, USA}
\author{P.V. Sushko}
\affiliation{Physical and Computational Sciences Directorate, Pacific Northwest National Laboratory, Richland, Washington 99352, USA}
\author{Andreas D. Wieck}
\author{Arne Ludwig}
\affiliation{Lehrstuhl f\"ur Angewandte Festk\"orperphysik, Ruhr-Universit\"at Bochum, D-44870 Bochum, Germany}
\author{Kai-Mei C. Fu}
\affiliation{Department of Physics, University of Washington, Seattle, Washington 98195, USA}
\affiliation{Department of Electrical Engineering, University of Washington, Seattle, Washington 98195, USA}

\begin{abstract}
Two-dimensional stacking fault defects embedded in a bulk crystal can provide a homogeneous trapping potential for carriers and excitons.  Here we utilize state-of-the-art structural imaging coupled with density functional and effective-mass theory to build a microscopic model of the stacking-fault exciton.  The diamagnetic shift and exciton dipole moment at different magnetic fields are calculated and compared with the experimental photoluminescence of excitons bound to a single stacking fault in GaAs.  The model is used to further provide insight into the properties of excitons bound to the double-well potential formed by stacking fault pairs.  This microscopic exciton model can be used as an input into models which include exciton-exciton interactions to determine the excitonic phases accessible in this system.

\end{abstract}

\date{\today}

\maketitle

\section{Introduction}

The stacking fault (SF), a misordering of lattice planes in a crystal lattice, is a prevalent two-dimensional (2D) crystal defect which can affect the mechanical, optical, and electrical properties of a material~\cite{ref:jamaati2014esf,ref:yang2002ees,ref:caroff2011cpn,ref:guha1993dbb}. While typically the macroscopic properties of a material are studied as a function of defect density~\cite{ref:colli2003cns}, the recent isolation of large-scale ($\sim$10~\textmu m) stacking faults in GaAs enabled the study of excitons bound to a single stacking fault~\cite{ref:karin2016gpd}. The high-homogeneity of the excitonic emission, combined with the measured giant static dipole moment, indicate the atomically-thin stacking-fault potential may be a promising platform for the realization of novel excitonic phases~\cite{ref:butov2002tbe,ref:butov2002mos}.
Due to the built-in static dipole moment, the excitons bound to the SFs in GaAs demonstrate the magneto-Stark effect: non-reciprocal variation of the exciton energy in a magnetic field~\cite{ref:karin2016gpd}. As shown in earlier studies of excitons in bulk materials, this effect provides a direct proof of exciton motion in the crystal~\cite{PhysRevLett.5.505,gross61}. It is also of importance in nonlinear optics in semiconductors, providing a mechanism of, e.g., second harmonic generation on otherwise forbidden excitonic states~\cite{PhysRevLett.110.116402}.

To gain a further insight into the magneto-optics of excitons, their lifetime and exciton-exciton interactions, knowledge of the confinement potential and wave function of the stacking-fault exciton is required. Advancements in structural imaging and density functional theory calculations, combined with our ability to optically isolate and characterize excitons on a single fault, provide an unprecedented opportunity to quantitiatively understand the stacking-fault exciton. As a result, in this paper we develop a microscopic model of the stacking-fault potential and SF exciton wave function in GaAs. Within this model the exciton hole is localized at the SF plane and the electron is bound via Coulombic attraction to the hole. An electric field due to the spontaneous polarization across a single SF plane is modelled by a step-function which results in the large electron-hole separation of about 10 nm. Variational method calculations based on this potential are found to be in reasonable agreement with experiment with respect to the observed diamagnetic shift and static dipole moment in single stacking faults. The model further provides an explanation for the two-fold larger dipole moment observed in double-well potentials formed by stacking fault pairs, suggesting that these double-well structures could provide further tunability in the excitonic properties.

The paper is organized as follows. In Sec.~\ref{sec:struct} we present the structural images of the SFs via electron microscopy. Further, we present the microscopic model of the SF potential in Sec.~\ref{sec:model}. Section~\ref{sec:compare2exp} provides a detailed comparison between the calculated excitonic properties of SFs with the experiment in terms of key parameters such as diamagnetic shifts and magneto-Stark effect demonstrating the validity of the model. The paper is summarized with a brief conclusion in Sec.~\ref{sec:concl}.

\section{Structural Imaging of Single and Double Stacking Faults}\label{sec:struct}

\begin{figure*}[!htb]
  \centering
  \includegraphics[width=6.5 in]{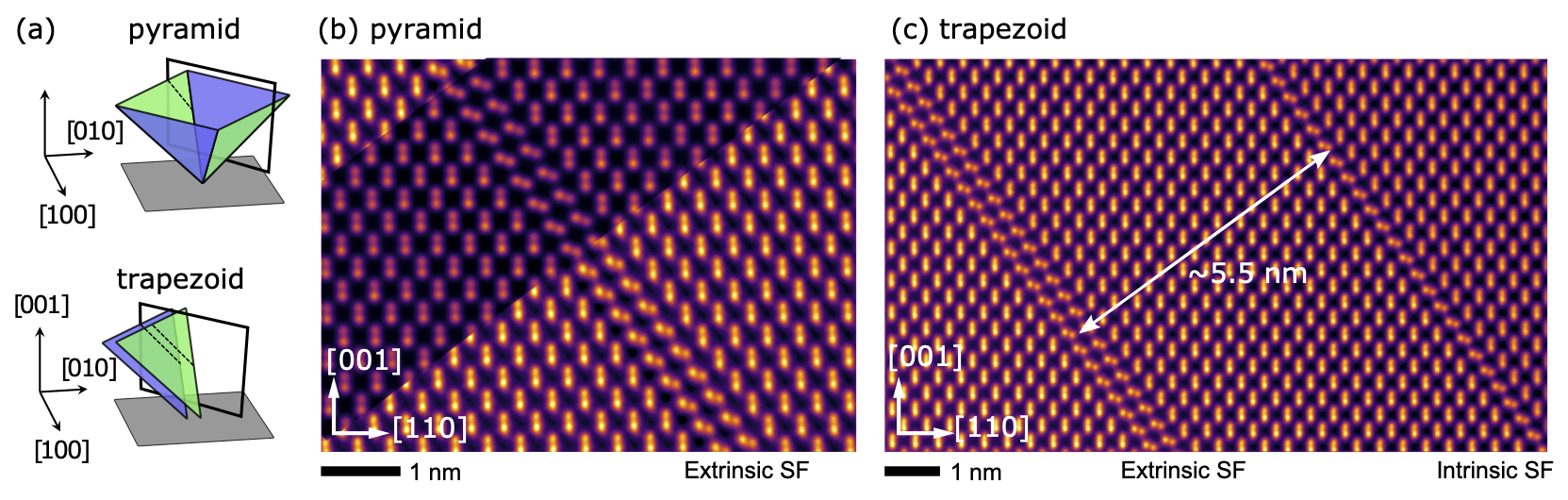}
  \caption{\label{fig:haadf} (a) 
  The geometry of the cross section of the STEM images. The black rectangles show the cross section plane and the dashed lines show where the stacking faults are. (b-c) Cross-sectional STEM images of stacking faults in the pyramid and trapezoid samples. (b) is overlaid with multislice image simulations based on \textit{ab initio} models, showing an excellent match between experiment and theory. These images are the result of non-rigid alignment and template matching, as described in Appendix \ref{sec:AppA}.}
\end{figure*}

Cross-sectional scanning transmission electron microscopy (STEM) analysis of two different stacking-fault defects, the pyramid and trapezoid, was performed to determine the structure of the defects. The experimental image is compared to the result of multislice image simulations based on \textit{ab initio} calculations (Appendix~\ref{sec:AppA}); these results show an excellent agreement. The stacking faults are embedded in the GaAs epitaxial layer which is grown on a (100)-terminated GaAs substrate. The location of the faults are identified by oval defects at the surface~\cite{ref:Kasai1998mod}. The geometry of the cross-section with respect to the structure is shown in the insets of Fig.~\ref{fig:haadf}(a).  As shown in Fig.~\ref{fig:haadf}(b), in the pyramid structure we observe an isolated stacking fault plane parallel to the (111) plane. Based on these observations, it is confirmed that excitons are bound to a single, highly homogeneous stacking-fault in the pyramid structure. In contrast, the trapezoid structure shown in Fig.~\ref{fig:haadf}(c) exhibits closely spaced intrinsic-extrinsic stacking-fault pairs. In this particular trapezoid, the planes are separated by $\sim5.5$~nm in the [111]-type direction, but this distance can vary from one structure to another. Thus for trapezoid structures, excitons are bound to a pair potential, in which the stacking-fault separation is expected to impact the bound-exciton properties.

\section{Microscopic model}\label{sec:model}

\subsection{Single stacking fault at zero magnetic field}\label{sec:zeroB}

We first consider excitons bound to a single stacking fault at zero magnetic field.  The stacking fault is positioned at $z = 0$, with $z\parallel [111]$, and occupies the $xy$-plane  with $x \parallel [11\bar{2}]$, $y \parallel [\bar{1}10]$. We consider excitons described by the wave function $\Psi_X = \psi(\bm r_e, \bm r_h) u_c(\bm r_e) u_v (\bm r_h)$, where $u_c$ and $u_v$ are the Bloch functions of the conduction band ($\Gamma_6$ representation of the $T_d$ point group) and the heavy-hole valence subband ($\Gamma_8, \pm 3/2$ representation of the $T_d$ point group), respectively, and $\psi(\bm r_e, \bm r_h)$ is the two-particle envelope function.

To obtain the exciton spectrum in the absence of a magnetic field, we solve the Schr\"odinger equation 
\begin{equation}
    \label{Schroed}
   \mathcal H \psi = \varepsilon \psi,
\end{equation} 
for the exciton envelope function $\psi(\bm r_e, \bm r_h)$ and energy $\varepsilon$ with the following Hamiltonian
\begin{multline}
\label{ham0}
\mathcal H = \frac{\bm p_e^2}{2 m_e} + \frac{p_{hx}^2 + p_{hy}^2}{2 m_{h,\parallel}} + \frac{p_{hz}^2}{2 m_{h,\perp}} + \\  + E_g + V_{\rm SF}(z_e, z_h) -\frac{e^2}{\varkappa|\bm r_e - \bm r_h|}\:.
\end{multline}
Here $\bm p_{e,h} = - \mathrm{i} \hbar \bm \nabla_{e, h}$ are the electron and hole momentum operators, $m_e$, $m_{h,\parallel}$ and $m_{h,\perp}$ are the components of the electron and hole effective-mass tensors, $E_g$ is the energy gap of the bulk material, $V_{\rm SF}$ is the stacking-fault potential experienced by the electron and hole, $e$ is the electron charge and $\varkappa$ is the static dielectric constant of the background medium. The electron effective mass is isotropic, whereas the heavy-hole effective-mass tensor has different components for the motion in the stacking fault plane ($m_{h,\parallel}$) and in the $z$-direction ($m_{h,\perp}$)~\cite{ivchenkopikus,ivchenko05a,ref:ikonic1992vss}. 

\begin{figure}[!htb]
  \centering
  \includegraphics[draft=false,width = 3.5 in]{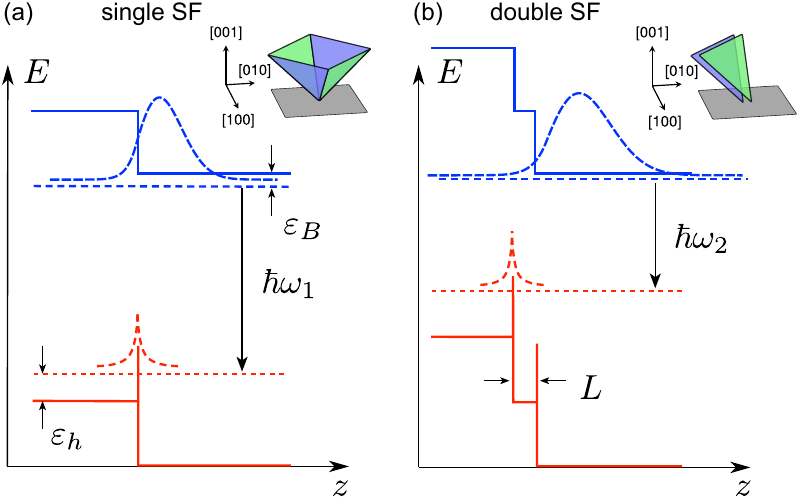}
  \caption{\label{fig:potential} Sketch of the conduction and valence band potentials for single (a) and double (b) stacking fault structures (in the electron representation). The blue and red dashed lines schematically depict the $z$-distribution of electron and hole density in the exciton, respectively.
$\varepsilon_h$ denotes the hole binding energy and $\varepsilon_B$ denotes the exciton binding energy. The insets illustrate the pyramid and trapezoid structure embedded in the crystal, $z$ is in the direction perpendicular to the stacking fault plane. 
  }

\end{figure}

We suggest that the presence of a single stacking fault modifies the electron and hole bands, yielding a potential in the following form
\begin{equation}
\label{pot}
V_{\rm SF}(z_e, z_h) = V_0 \Theta(-z_e) - V_0 \Theta(-z_h) - u_0 \delta(z_h)\:,
\end{equation}
where $\Theta(z)$ and $\delta(z)$ are the Heaviside and Dirac delta functions, respectively. $V_0$ and $u_0$ are positive parameters. This potential is sketched in Fig.~\ref{fig:potential}(a). The model potential binds the hole in $z$-direction due to the $\delta$-function term, but does not bind the electron. The electron in the exciton is then attracted to the hole due to the Coulomb interaction.
The parameter $V_0$ describes the band offset related to the presence of the built-in spontaneous electric polarization, and consequently the electric field, in the stacking fault layer~\cite{PhysRevB.86.081302, PhysRevB.87.035305, doi:10.1063/1.4880209}. Thus, $V_0$ is equal to the electrostatic potential change across the SF. The delta-function term that confines the hole models the type-II band alignment, which is believed to appear between the GaAs zinc-blende and wurtzite phases~\cite{ref:spirkoska2009sop,ref:belabbes2012ebt,ref:heiss2011dcc}. The same term for the conduction band is neglected since it does not bind an electron and, hence, only slightly modifies electron wave function. The suggested potential agrees well with density functional theory (DFT) calculations of the stacking fault electrostatic potential and single-particle wave functions which predicts $V_0\approx 10$~meV and a hole confinement length of $\sim 4$~nm (see Appendix~\ref{appendix:DFT} for details). 

The confinement energy of a hole bound to potential~\eqref{pot} is 
\begin{equation}
    \label{hole_singleSF}
    \varepsilon_h = \varepsilon_0 \left( 1 - \frac{V_0}{4 \varepsilon_0} \right)^2\:,
\end{equation}
where $\varepsilon_0 = m_{h,\perp} u_0^2/(2\hbar^2)$. The potential binds the hole if $V_0 < 4 \varepsilon_0$, which is true for our system, where $V_0 \approx 10$~meV and $\varepsilon_h \approx 10$~meV (corresponding to $\varepsilon_0 \approx 15$~meV), as will be shown below. The localization length of the heavy hole in the $z$-direction is $a_h \sim [2\hbar^2/(m_{h,\perp} \varepsilon_h)]^{1/2}$. To simplify the calculation, in the following we assume that $a_h = 0$, so that the hole is tightly bound to the stacking fault and $z_h = 0$. The validity of this assumption is supported by the $\sim$4~nm DFT hole confinement length, which is much less than the $\sim 20$~nm  exciton diameter. By contrast, the electron remains bound only due to the Coulomb interaction with the hole. Note, that the model potential of ZnSe SFs suggested recently in~\cite{ref:smirnov2018ebi} does not bind a hole. It may be related to large electric field inside the ZnSe SFs as compared to GaAs SFs ($\sim 5$ times larger), which prevents the binding of a hole (as described by Eq.~\eqref{hole_singleSF} at $V_0 >4 \varepsilon_0$).

In the absence of an external magnetic field, the exciton envelope can be written as $\psi(\bm r_e, \bm r_h) = \varphi(\bm r) {\exp{(\mathrm{i} \bm K \bm R)}}$, where $\bm r = \bm r_e - \bm r_h$ is the coordinate of relative motion {(note, that ${z} = z_e$)}, and $\bm R$ and $\bm K$ are the coordinate and the wave vector of the exciton center-of-mass in the stacking fault plane. The effective Hamiltonian that acts on the exciton envelope function $\varphi(\bm r)$ is
\begin{equation}
\label{ham1}
\mathcal H_0 = \frac{p_x^2 + p_y^2}{2 \mu} + \frac{p_{z}^2}{2 m_e} + V_0 \Theta(-z) - \frac{e^2}{\varkappa |\bm r|}\:,
\end{equation}
with $\mu^{-1} = m_e^{-1} + m_{h,\parallel}^{-1}$.

To solve the Schr\"odinger equation with the Hamiltonian~\eqref{ham1} we use the variational approach. We choose $\varphi(\bm r)$ in the form 
\begin{equation}
\label{trial}
\varphi(\bm r) = \mathcal{N} \exp \left( - \sqrt{\frac{\rho^2}{a^2} + \frac{z^2}{c^2}} \right) f\left( \frac{z}{c} \right)\:,
\end{equation}
where $\bm \rho = (x,y)$, $\mathcal N$ is the normalization constant,
\begin{equation}
\label{trial_z}
f(\xi) = (1+\alpha \xi) \Theta(\xi) + \mathrm{e}^{\alpha \xi} \Theta(-\xi)\:,
\end{equation}
and $a$, $c$ and $\alpha$ are variational parameters.
The parameters $a$ and $c$ are the effective in-plane and $z$-sizes of the exciton, the function $f(\xi)$ describes the asymmetric confinement of the electron. Correspondingly,  the dimensionless parameter $\alpha > 0$ determines the asymmetry of the exciton wave function in the $z$-direction, which is caused by the asymmetry of the electron distribution. The wave function~\eqref{trial_z} well describes the behavior of the electron $z$-distribution with the change of $V_0$ in Eq.~\eqref{pot}: At $\alpha = 0$, which corresponds to $V_0 = 0$, we have a symmetric distribution, $f(\xi) = 1$, and at $\alpha \gg 1$, which corresponds to large values of $V_0$, the wave function $f(\xi)$ vanishes at $\xi \leq 0$, and the electron does not penetrate the barrier. 

\begin{figure}[!htb]
  \centering
  \includegraphics[draft=false,width = 3.5 in]{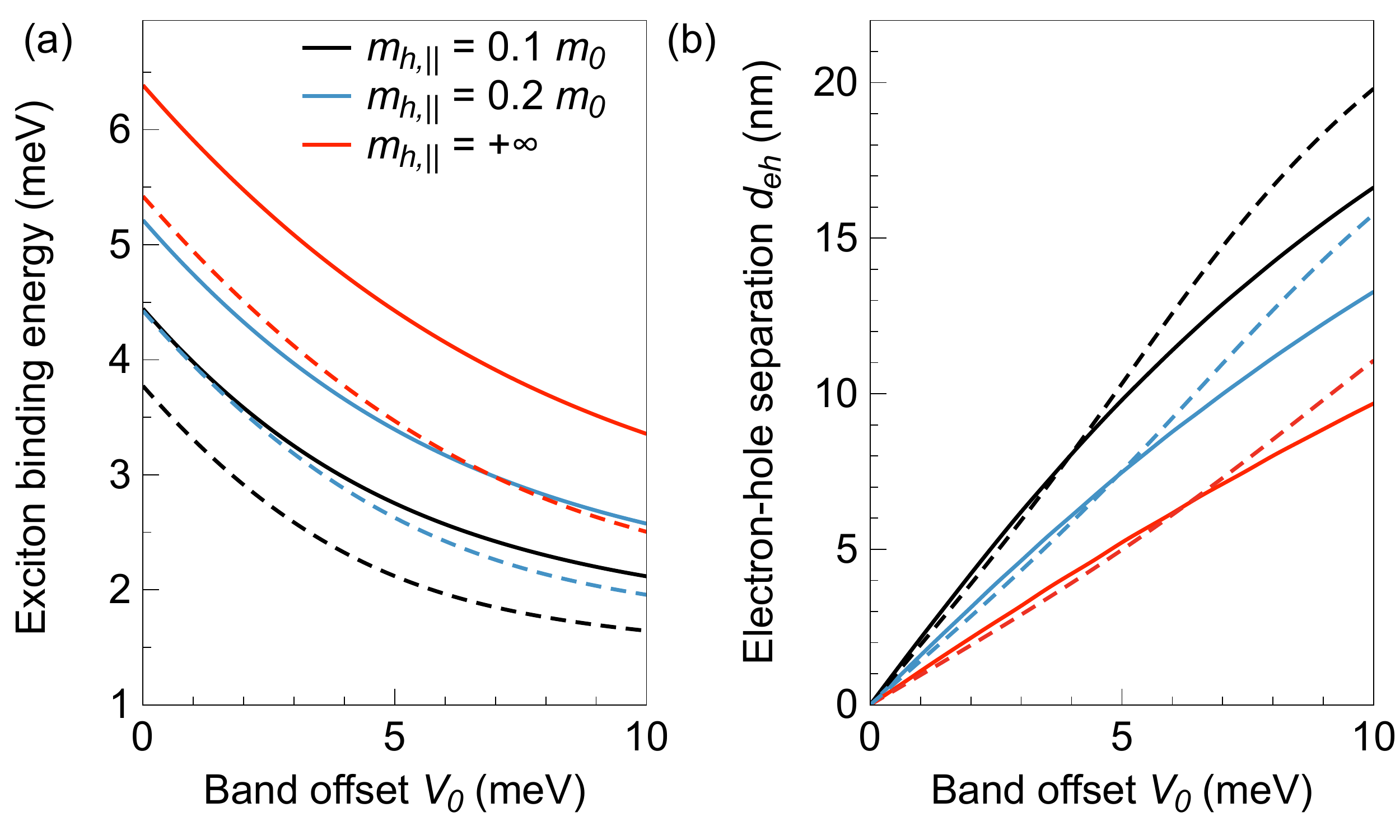}
  \caption{\label{fig:zero_field} Exciton binding energy (a) and electron-hole separation (b) as functions of the stacking fault band offset. Solid and dashed lines depict the results for the hydrogen-like and Gaussian trial exciton wave functions.   }
\end{figure}

Figure~\ref{fig:zero_field} shows the results of our variational calculations for the Hamiltonian~\eqref{ham1} and the trial function~\eqref{trial}. We plot the  exciton binding energy $\varepsilon_B = - \langle \varphi|\mathcal H_0|\varphi\rangle$ and the average distance between the electron and hole in the $z$-direction $d_{eh} = \langle \varphi|z|\varphi \rangle$, as functions of the band offset $V_0$.  In the calculations we use $m_e = 0.07~m_0$, $\varkappa = 12$, and three different values for $m_{h,\parallel}$: $0.1~m_0$, $0.2~m_0$ and $+\infty$. The infinite case corresponds to an electron bound on a donor that is located at the stacking fault plane. 

Additionally, the dashed lines in Fig.~\ref{fig:zero_field}(a) present the results using a simplified Gaussian-like trial wave function
\begin{equation}
\label{wfgauss}
\tilde{\varphi}(\bm r) = \mathcal{N} \exp \left( - \frac{\rho^2}{2a^2} - \frac{z^2}{2c^2} \right) f\left( \frac{z}{c} \right)\:,
\end{equation}
with $f$ given by Eq.~\eqref{trial_z}. It is known that the Gaussian-like function underestimates the binding energy of the Coulomb potential by 25\%. Also by comparing solid and dashed lines in Fig.~\ref{fig:zero_field}a we conclude that the trial wave function~\eqref{wfgauss} results in a 20$-$30 \% smaller exciton binding energy than the more accurate hydrogen-like wave function~\eqref{trial}. However, the results for the electron-hole separation using the Gaussian-like trial function agree well with the ones 
obtained for the hydrogen-like trial function. This agreement motivates using a Gaussian-like trial wave function to calculate the electron-hole separation in the more complicated case where the magnetic field $\bm B \neq$ 0.

\subsection{Single stacking fault at non-zero magnetic field}
\label{sec:nzm}
An external magnetic field $\bm B$ applied in the stacking fault plane brings the electron and hole closer and shrinks the exciton wave function. This results in the diamagnetic shift, which is quadratic in $\bm B$, and also in the decrease of $d_{eh}$, yielding the suppression of the magneto-Stark effect. In a wide range of magnetic fields applied in the experiment, these effects cannot be treated perturbatively.  Thus, we now consider an exciton bound at the single stacking fault in the presence of an external magnetic field $\bm B \parallel y$. The exciton Hamiltonian is obtained from Eq.~\eqref{ham0} using the substitution $\bm p_e \to \bm p_e - (e/c) \bm A(\bm r_e)$ and $\bm p_h \to \bm p_h + (e/c) \bm A(\bm r_h)$, where $\bm A$ is the vector potential chosen to be the symmetric form
$\bm A(\bm r) = B(z/2,0,- x/2)$.

Since we assume the strong hole confinement in the $z$-direction, we can neglect the influence of the in-plane magnetic field on the heavy-hole motion along the stacking fault normal. Hence, the exciton diamagnetic shift including the field-induced variation of the average electron-hole separation is determined by the electron component. In the presence of a magnetic field $\bm B \parallel y$, the momentum of the exciton center of mass should be written as
\begin{equation}
\label{COM}
P_x = -\mathrm i \hbar \frac{\partial}{\partial x_e} -\mathrm i \hbar \frac{\partial}{\partial x_h} %p_{ex} + p_{hx} 
- \frac12 m_e \omega_c z\:,~~~ P_y = p_{ey} + p_{hy}\:,
\end{equation}
where $\omega_c = |e|B/(m_e c)$~\cite{ref:gorkov1968ctm}, see also Refs.~\cite{PhysRevB.76.075303,0268-1242-13-3-007,PhysRevB.63.153307} in which quasi-two dimensional excitons in an in-plane magnetic field were studied\addMisha{.} The exciton envelope wave function then reads
\begin{equation}
\label{Xwf_B}
\psi(\bm r, \bm R) = \exp \left[ \mathrm{i} \left( P_x + \frac12 m_e \omega_c z \right) \frac{X}{\hbar} + \mathrm{i} P_y \frac{Y}{\hbar}  \right] \varphi(\bm r)\:.
\end{equation}
Here $P_x$ and $P_y$ are the eigenvalues of the center of mass momentum operator in Eq.~\eqref{COM}.

Using the wave function~\eqref{Xwf_B} and the general Hamiltonian~\eqref{ham0}, we obtain the effective Hamiltonian that describes the internal motion of the exciton (at $P_x=P_y=0$):
\begin{multline}
\label{HB}
\mathcal H_{B} = \frac{[p_x + (m_e - \mu/2) \omega_c z]^2}{2 m_e} + \frac{(p_z -  \mu \omega_c x/2)^2}{2m_e} + \\ + \frac{(p_x - \mu \omega_c z/2)^2}{2m_{h,\parallel}} + \frac{p_y^2}{2\mu} + V_0 \Theta(-z) - \frac{e^2}{\varkappa |\bm r|}\:.
\end{multline}
To obtain the ground state of the exciton in a magnetic field, we use the following trial wave function
\begin{equation}
\label{wfB}
\varphi_B(\bm r) = \mathcal N \exp \left( - \frac{x^2}{2a^2} - \frac{y^2}{2b^2} - \frac{z^2}{2c^2} \right) f\left( \frac{z}{c} \right)\:
\end{equation}
with four variational parameters $a$, $b$, $c$ and $\alpha$, and $f$ given by Eq.~\eqref{trial_z}. At $B = 0$, we have $a = b$ and this wave function coincides with~\eqref{wfgauss}. Although this wave function does not allow one to evaluate accurately the exciton binding energy at $B=0$, it provides reasonable accuracy for the electron-hole separation and allows to substantially simplify numerical calculations as discussed above. 
The diamagnetic shift of the exciton energy is then determined by
\begin{equation}
\label{eq:dia}
    E_D = \langle \varphi_B | \mathcal H_B | \varphi_B \rangle - E_0\:,
\end{equation}
where $E_0 = \langle \varphi_B | \mathcal H_B | \varphi_B \rangle$ at $B = 0$.

\begin{figure}[tb]
  \centering
  \includegraphics[draft=false,width = 3.7 in]{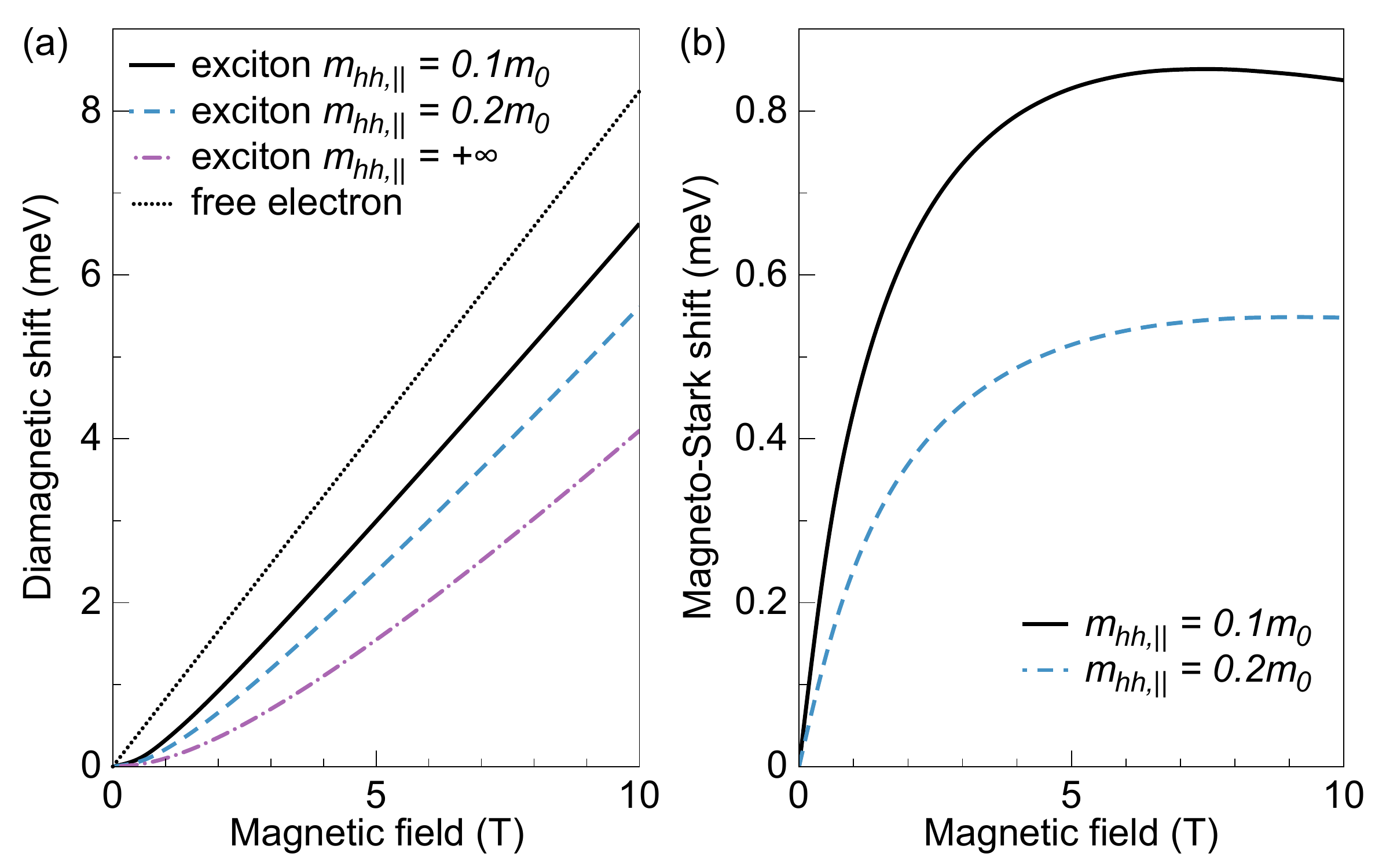}
  \caption{\label{fig:dia} Calculated exciton diamagnetic shift (a) and magneto-Stark shift (b) for different values of the hole in-plane mass $m_{hh, \parallel}$. The dotted line in panel (a) is the diamagnetic shift of a free electron ($\hbar \omega_c/2$). Here, $V_0 = 10$~meV.
  }
\end{figure}

In order to evaluate the magneto-Stark effect, we calculate the center of mass dispersion of the exciton making use of the following relations for the exciton velocity $\bm v(\bm P)$ and the exciton kinetic energy $E(\bm P)$~\cite{ref:gorkov1968ctm}:
\begin{equation}
\label{disper}
\bm v = \frac{dE(\bm P)}{d\bm P}, \quad \bm P = M \bm v(\bm P) - \frac{e}{c} [\bm B \times \bm r].
\end{equation}
where $M=m_e+ m_{h,\parallel}$ is the mass for exciton translational motion in the SF plane.
Solving Eq.~\eqref{disper}, we obtain
\begin{equation}
\label{disper1}
E(\bm P) = \frac{(P_x - e B d_{eh}/c)^2}{2M} + \frac{P_y^2}{2M}.
\end{equation}
Here the electron-hole separation $d_{eh}$ generally depends on the magnetic field. Equation~\eqref{disper1} allows us to evaluate the magneto-Stark shift of the exciton energy as
\begin{equation}
\label{MS}
E_{S} = - \frac{e P_x B d_{eh}}{Mc} = \beta' K_x B,
\end{equation}
where the parameter $\beta' = - e \hbar d_{eh}/(Mc)$ describes the slope of the magneto-Stark shift and $K_x = P_x/\hbar$ is the $x$-component of the exciton wave vector that is defined by the experiment geometry, see Eq.~\eqref{eq:inplanep} in Sec.~\ref{sec:compare2exp}. At low magnetic fields $\beta'$ does not depend on the magnetic field and is determined by the electron-hole separation $d_{eh}$ at $B = 0$, which is calculated in Sec.~\ref{sec:zeroB} and shown in Fig.~\ref{fig:zero_field}b.

Figure~\ref{fig:dia} illustrates the dependence of the diamagnetic shift~\eqref{eq:dia} and magneto-Stark shift~\eqref{MS} on the magnetic field. The magnetic field lying at the SF plane shrinks the exciton wave function in the $xz$-plane and, thus, reduces the electron-hole separation $d_{eh}$. Therefore, magneto-Stark shift grows sublinearly with increasing $B$, tends to saturation at large fields and then decreases at even larger fields, when the reduction of $d_{eh}$ is faster than $\propto 1/B$. The diamagnetic shift dependence changes from quadratic to linear in $B$ with increasing magnetic field. At large $B$ the exciton diamagnetic shift is equal to the one of a free electron modified by logarithmic corrections due to effective one-dimensional Coulomb attraction to the hole~\cite{landau3eng}.

Here we provide a brief comparison of the theoretical and experimental magneto-Stark slope $\beta'K_x$ at low magnetic fields to illustrate that the model is reasonable. A comparison to the full experimental field dependence of the diamagnetic and magneto-Stark shifts will be given in Sec.~\ref{sec:compare2exp}. The experimental value of the parameter $\beta'_{exp}K_x \approx 350$~$\mu$eV/T measured in the pyramid stacking fault~\cite{ref:karin2016gpd} corresponds to $\beta'_{th}K_x$ for a reasonable set of values, i.e.,  $m_{h,\parallel} = 0.1~m_0$, $V_0 = 7$~meV using the experimental value of $|K_x| \approx 1.6\times10^5$~cm$^{-1}$. The choice of parameters is not unique.  For example, the same value of the magneto-Stark slope can be achieved at  $m_{h,\parallel} = 0.14~m_0$ and $V_0 = 10
$~meV. This ambiguity is related to the fact that the exciton mass and the electron-hole distance enter only as a combination $d_{eh}/M$, thus, simultaneous increase of the exciton mass and electron-hole separation (by increasing the band offset $V_0$, see Fig.~\ref{fig:zero_field}(b)) results in the same value of $\beta'$ in Eq.~\eqref{MS}. The value of the electric field inside the SF that corresponds to $V_0 = 10$~meV and the width of SF $~10$~\AA~(see Appendix~\ref{appendix:DFT} for details) is $F \approx 0.1$~MV/cm, which is in line with the experiments on polytypic GaAs nanowires, where $F$ lies in the range of 0.18 to 0.27 MV/cm~\cite{doi:10.1063/1.4880209}. On the other hand, this electric field is about 5 times smaller than in ZnSe SFs~\cite{ref:smirnov2018ebi} and about 25 times smaller than in GaN SFs~\cite{PhysRevB.86.081302}.

\subsection{Double stacking fault}
\label{sec:dsf}

Besides the pyramid configuration, when SF planes are isolated, SFs can appear in a form of closely lying parallel planes, Fig.~\ref{fig:haadf}(b). In this trapezoid configuration an exciton is bound to a double SF potential sketched in Fig.~\ref{fig:potential}(b). We model this potential as a sum of two single SF potentials with the same band offset:
\begin{equation}
    \label{eq:doubleV}
    V_{\rm 2 SF} (z_e, z_h) = V_{\rm SF}(z_e, z_h) + V_{\rm SF}(z_e - L, z_h - L)\:,
\end{equation}
where $L$ is a separation between SF planes, and $V_{\rm SF}$ is a single SF potential given by Eq.~\eqref{pot}. The assumption that both SFs in a pair have the same direction of the built-in electric field follows from the experimentally observed approximately twice increase of the exciton electric dipole moment as compared to the single SF case (see Sec.~\ref{sec:trap_exp} for details) and the DFT calculations (see Appendix~\ref{appendix:DFT}).

\begin{figure}[!htb]
  \centering
  \includegraphics[draft=false,width = 2 in]{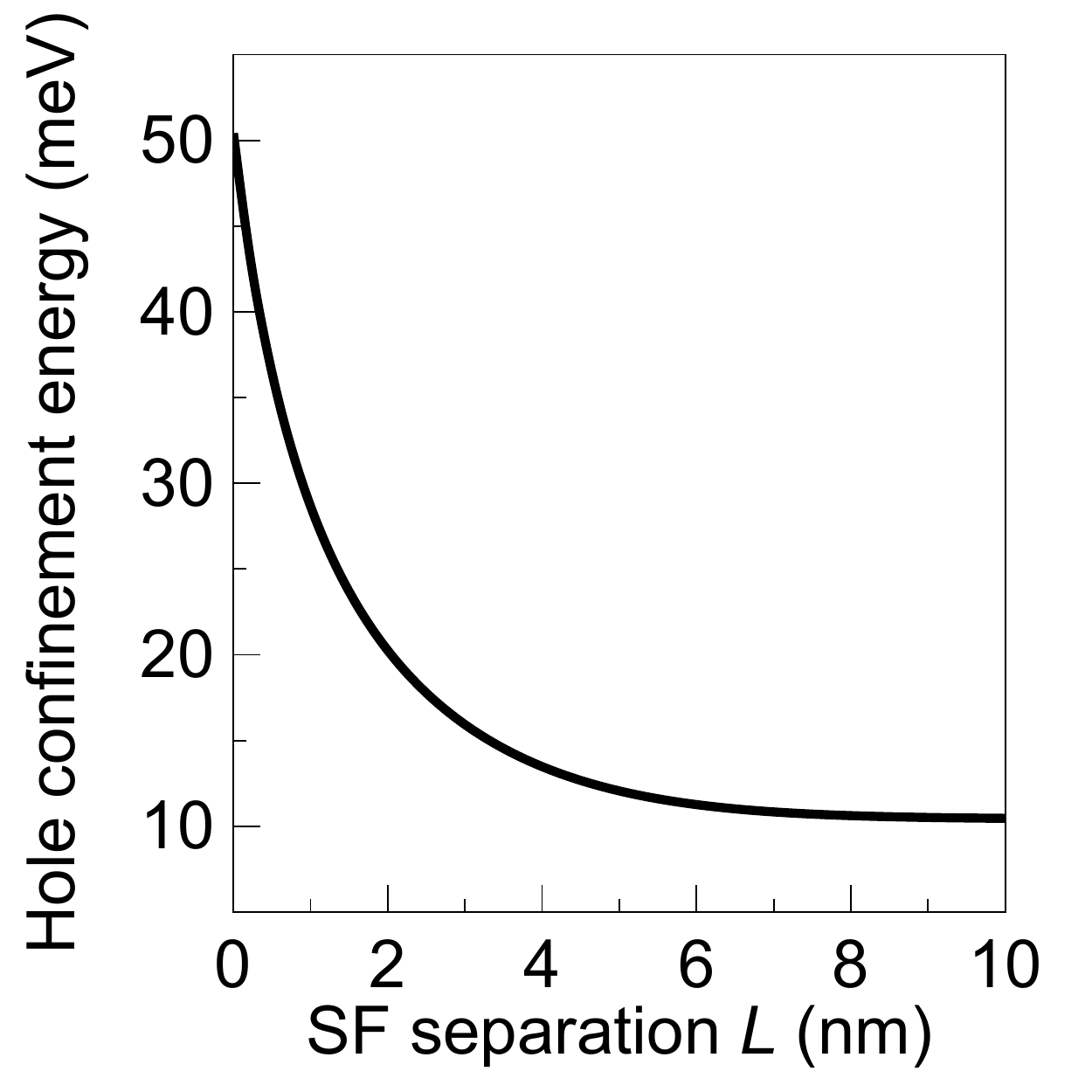}
  \caption{\label{fig:hole2SF} Hole confinement energy as a function of SF separation.
  $V_0 = 10$~meV, $\varepsilon_0 = 15$~meV.} 
\end{figure}

We assume that the separation between the SFs is of the order of the hole confinement length in $z$-direction $a_h$, which is around a few nanometers, but is much smaller than the electron-hole separation $d_{eh}$, which is of the order of tens of nanometers. In that case an electron ``sees'' the double SF structure as a single SF with a twice increased built-in electric field (band offset equal to $2 V_0$), and hence, as it follows from Fig.~\ref{fig:zero_field}, the $d_{eh}$ parameter for a double SF also increases approximately two fold. On the other hand, the hole energy depends significantly on the SF separation. If $L \gg a_h$, the hole resides at the SF at $z_h = 0$ and does not ``feel'' another SF. With the decrease of $L$, when $L \sim a_h$, the hole confinement energy increases and its wave function is distributed over both SFs. In the limit $L = 0$ the hole energy is found from Eq~\eqref{hole_singleSF} with $u_0 \to 2u_0$ and $V_0 \to 2 V_0$, respectively. The dependence of hole confinement energy $\varepsilon_h$ on $L$ is shown in Fig.~\ref{fig:hole2SF}.

The scheme of exciton optical recombination is sketched in Fig.~\ref{fig:potential}. The transition energies of an exciton bound to a single and double SFs are
\begin{eqnarray}
    \label{hw}
    \hbar \omega_1 &=& E_g - V_0 - \varepsilon_{h1} - \varepsilon_{B1}\:, \nonumber \\
    \hbar \omega_2 &=& E_g - 2V_0 - \varepsilon_{h2} - \varepsilon_{B2}\:,
\end{eqnarray}
where $\varepsilon_{h1(2)}$ is the confinement energy of a hole bound to a single (double) SF potential, and $\varepsilon_{B1(2)}$ is the corresponding exciton binding energy.
Neglecting the difference between SF and bulk exciton binding energies, the shifts of SF-bound exciton PL lines with respect to the bulk one are $\hbar \omega_X^{3D} - \hbar \omega_1 \approx V_0 + \varepsilon_{h1}$ and $\hbar \omega_X^{3D} - \hbar \omega_2 \approx 2V_0 + \varepsilon_{h2}$.

\begin{figure}[!htb]
  \centering
  \includegraphics[draft=false]{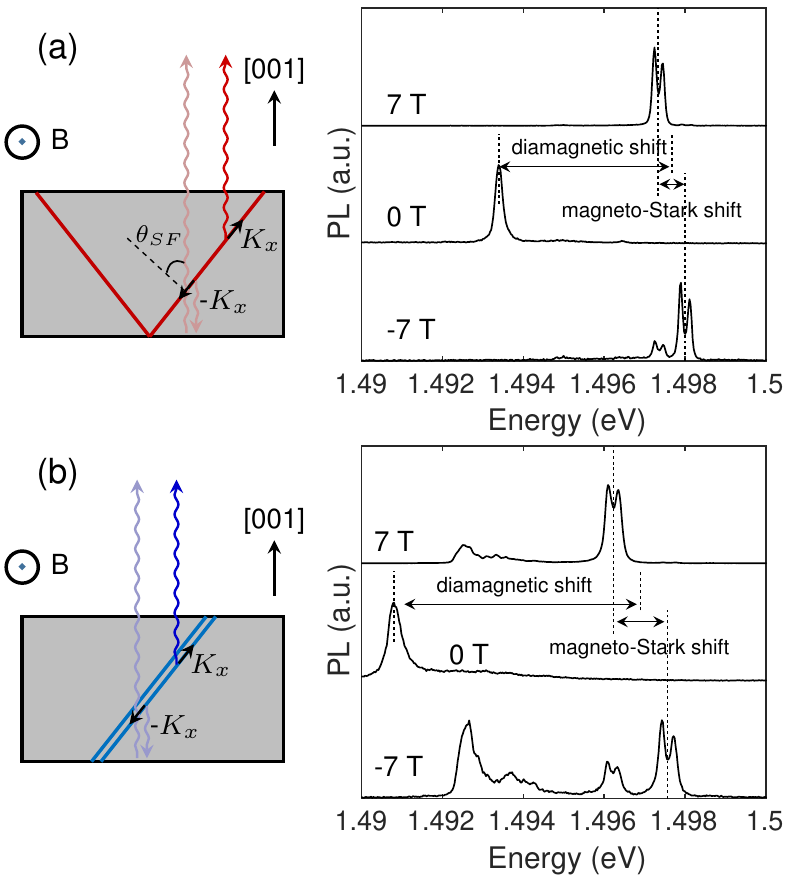}
  \caption{\label{fig:magnetoPL} Experimental geometry and PL spectrum of the (a) pyramid and (b) trapezoid structure. PL collected from $-K_x$ exciton is much weaker than PL from $+K_x$ exciton. The temperature is 1.5~K. The excitation laser is at 810~nm (1.53 eV).}
\end{figure}

\section{Comparison to experiment}
\label{sec:compare2exp}
\subsection{Magneto-photoluminescence}

Photoluminescence (PL) spectra taken at different magnetic fields are studied to verify the microscopic model. The MBE-grown GaAs sample~\cite{ref:karin2016gpd} is mounted in a continuous helium flow cryostat at 1.5~K with a variable magnetic field from 0 to 7~T. PL from both pyramid and trapezoid 10~$\mu$m-scale stacking fault structures is clearly resolved using an optical confocal setup with a resolution of $\sim$1~$\mu$m. The experimental geometry and typical PL spectra at $B = 0$ and $B = \pm7$~T are shown in Fig.~\ref{fig:magnetoPL}. The crystal [001] direction is perpendicular to the magnetic field $\bm B \parallel [\bar{1}10]$ and parallel to the optical axis. The collected stacking fault PL corresponds to excitons with in-plane momentum
\begin{equation} \label{eq:inplanep}
 K_x = \frac{\omega n}{c} \sin{\theta_{\mathrm{SF}}}
\end{equation}
where $\theta_{\mathrm{SF}}=54.7^{\circ}$ is the angle between the stacking fault normal and the emitted photon momentum, $\omega$ is the photon frequency, $n$ is the refractive index and $c$ is the speed of light. The collected PL from excitons with a wave vector $-K_x$ is much weaker than the PL from $+K_x$ excitons, because $-K_x$ excitons emit photons propagating towards the substrate and only backscattered light can be collected. %Thus, the collection angle of light determines the momentum of the excitons whose luminescence is measured. %Experimentally, the NA $= 0.7$ objective lens collects light from excitons with $K_x$ within 7 $\%$ of $K_{\mathrm{SF}}$.

\begin{figure}[!htb]
  \centering
  \includegraphics[draft=false,width=3.5in]{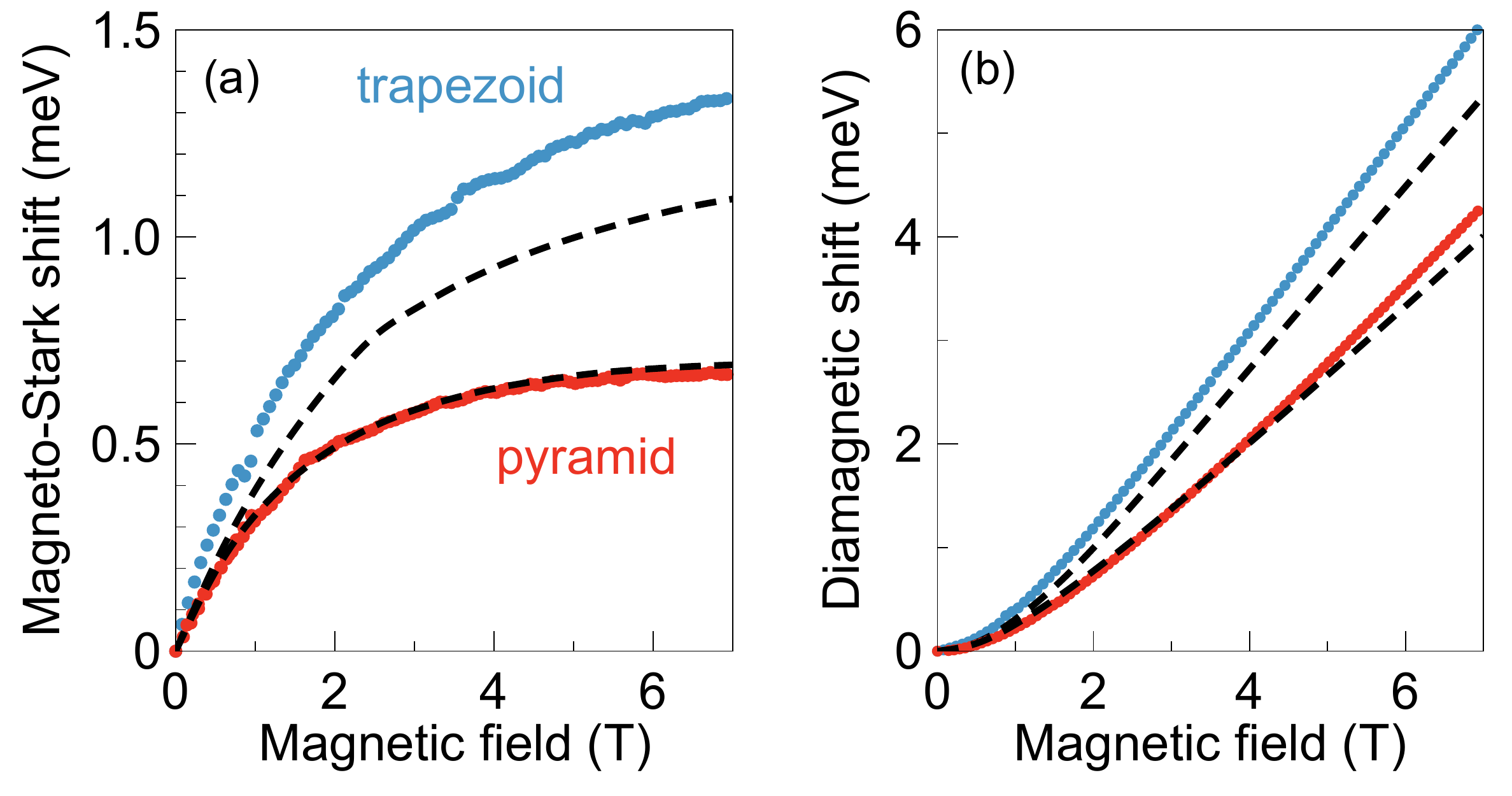}
  \caption{\label{fig:comparison} Comparison of experiment (points) and theory (dashed lines) for pyramid and trapezoid SFs. The parameters used in calculations are $m_{h,\parallel} = 0.14~m_0$, $V_0 = 10$~meV for pyramid SF, and $m_{h,\parallel} = 0.14~m_0$, $V_0 = 20$~meV for trapezoid SF, respectively. 
  }
\end{figure}

The magneto-PL spectra have similar properties for both the pyramid and trapezoid structures. At $B = 0$~T, a single PL peak is observed due to the recombination of excitons bound to the stacking fault plane. At $B = \pm7$~T, the main peak is split into a doublet due to the electron Zeeman splitting. At $B = -7$~T, in addition to the main doublet, a weaker doublet is observed at lower energy. This doublet has the same energy as the peak at $B = 7$~T, and thus,
is attributed to excitons with $-K_x$ momentum. The origin of the peaks near $\hbar \omega = 1.4925$~eV observed in the trapezoid structure at $B = \pm7$~T is unknown. %At high magnetic fields, the PL has a blue shift compared with PL at 0 field due to the diamagnetic shift. There is also a difference in emission energy at positive and negative magnetic field, which is due to the magneto-Stark effect. 
The diamagnetic and magneto-Stark shifts are clearly observed in the magneto-PL spectra, as illustrated in Fig.~\ref{fig:magnetoPL}.  Figure~\ref{fig:comparison} shows these experimental shifts as a function of magnetic field. The $B$-field dependence of diamagnetic and magneto-Stark shifts is in agreement with the microscopic model presented in Sec.~\ref{sec:nzm}, i.e. at low field the diamagnetic shift is $\propto B^2$ and the magneto-Stark shift is $\propto B$, whereas at high field diamagnetic shift tends to a linear $B$-dependence and the magneto-Stark shift exhibits a sublinear $B$-dependence. The origin of the change at high field is the decrease of the electron-hole separation $d_{eh}$ induced by the magnetic field. 

By fixing $m_{h,\parallel} = 0.14~m_0$ and using $V_0 = 10$~meV to fit the slope of the magneto-Stark shift at low magnetic fields, we obtain a qualitative agreement between the experimental and theoretical data for the pyramid structure at both low and high $B$-fields, as shown by the dashed lines in Fig. \ref{fig:comparison}. As suggested by the double SF model presented in Sec.~\ref{sec:dsf}, an electron in the trapezoid SF experiences the two-fold increase of the electric field as compared to the pyramid SF. By taking $m_{h,\parallel} = 0.14~m_0$ and $V_0 = 20$~meV, we obtain a reasonable agreement between the experimental and theoretical data for the trapezoid structure, see Fig. \ref{fig:comparison}. The larger value of the magneto-Stark shift in the experiment as compared to the theory might be caused by several reasons. One of the reasons is that the trapezoid structure consists of extrinsic and intrinsic SFs, see Fig.~\ref{fig:haadf}c, which may have different values of $V_0$. Thus, the actual increase of effective electric field in the double SF as compared to the single one might be larger than two. Another reason might be a slight increase of $d_{eh}$ with increased SF separation, which is not taken into account in the theory.

\begin{figure*}[!bht]
  \centering
    \includegraphics[draft=false,width = 6 in]{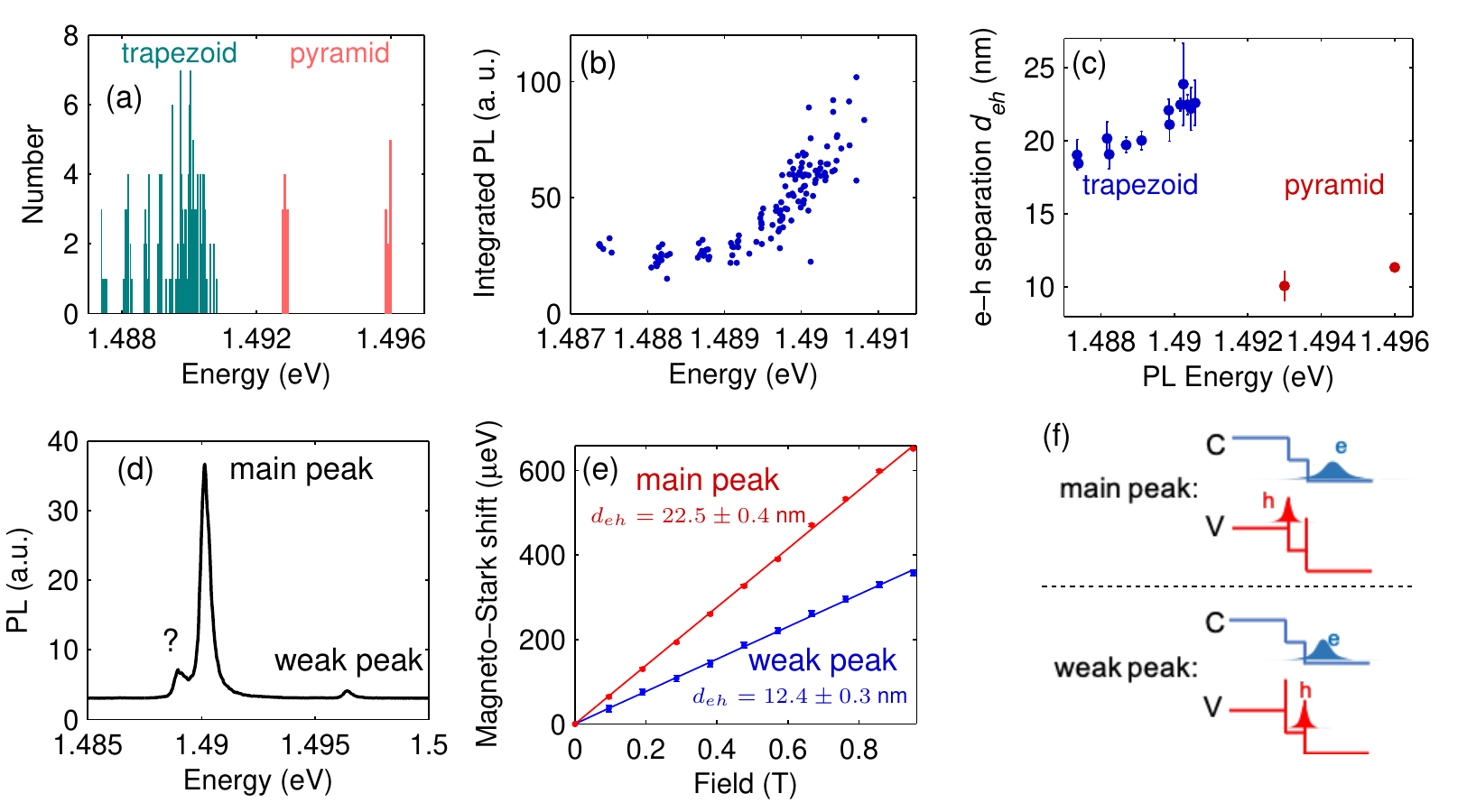}
  \caption{ \label{fig:trapezoid} (a) Distribution of the PL energy for excitons bound to the trapezoid and pyramid structures. 1.5 K. 810~nm excitation. (b) PL intensity as a function of $E_{\mathrm{SF}}$. (c) Electron-hole separation as a function of the $E_{\mathrm{SF}}$ for trapezoid and pyramid structures. The electron-hole separation is measured by the magneto-Stark effect and the calculation uses in-plane exciton mass 0.17$m_0$. (d) PL spectrum of a trapezoid with high PL energy. The peak marked by the question mark is unknown and does not exhibit magneto-Stark effect. (e) Magneto-Stark shift as a function of magnetic field for the main peak and 2nd peak. The calculated e-h separation $d_{eh}$ is listed in the figure. (f) Illustration of the exciton wave function for excitons of the two different peaks. C: condunction band. V: valence band. e: electron. h: hole.
  }
\end{figure*}

\subsection{Variance of PL in trapezoid structures} \label{sec:trap_exp}

The PL properties from different trapezoid structures exhibit a large variance relative to the pyramid structures. This is attributed to the variable separation between the two parallel stacking fault planes. Figure~\ref{fig:trapezoid}(a) shows the distribution of 0~T PL energies for 133 different trapezoid structures and 5 different pyramid structures (corresponding to 20 single stacking fault planes). For the pyramid structures, the stacking-fault excitons only emit at PL energy $E_{\mathrm {SF}} = 1.4928$ and 1.4959~eV. For the trapezoid structures, $E_{\mathrm {SF}}$ varies from 1.487 to 1.491~eV. Between 1.4875 and 1.4891~eV, 4 discrete energies are observed: 1.4875, 1.4882, 1.4887 and 1.4891~eV. At higher energies, the distribution is continuous.  

To further understand this effect, the trapezoid PL intensity and electron-hole separation are investigated, as shown in Fig.~\ref{fig:trapezoid}(b-c). The electron-hole separation is derived from the magneto-Stark shift, see Eq.~\eqref{MS}. Throughout this discussion we assume $M = 0.17 m_0$.  The electron-hole separation for the trapezoid is approximately double that of the pyramid and it slightly increases with $E_{\mathrm{SF}}$. The PL intensity for trapezoids emitting at one of the 4 discrete energies does not change with $E_{\mathrm{SF}}$. In contrast, for trapezoids with $E_{\mathrm{SF}}>$ 1.4891~eV, the intensity increases with increasing $E_{\mathrm{SF}}$. We further note that for these high energy trapezoids, the PL spectra contain two distinct stacking-fault exciton peaks, as shown in Fig.~\ref{fig:trapezoid}(d). The main peak corresponds to an exciton with electron-hole separation of 22.5~nm and the weak peak corresponds to an electron-hole separation of 12.4~nm. Possibly, this can be attributed to two different electron locations, as shown in Fig.~\ref{fig:trapezoid}(f). %\addMisha{\sout{The main emission peak corresponds to a hole bound to the 1st stacking fault and the electron repelled on the other side of the 2nd stacking fault. For the weak peak, the electron and hole are bound to the same fault.}} 
The quantitative analysis of the interplay between the main and the weak peaks is beyond the scope of this work.
%\mishad{\textit{So far this discussion contradicts the model described in Sec.~\ref{sec:dsf}.}} \kaimei{It is not clear to me why this contradicts IIIC.}  \addMisha{\textit{Let us check that it does not read as if the main effect is the increase in the distance between the SFs.}}
For trapezoids with PL energy at one of the 4 discrete values between 1.4875 to 1.4891~eV, the 2nd peak is not observed. We attribute this to a higher tunneling rate from the metastable (weak peak) configuration to the stable configuration (main peak) or a delocalization of the hole wavefunction over both faults (Sec.~\ref{sec:dsf} when the two stacking faults are close.)

It follows from the microscopic model shown in Sec.~\ref{sec:dsf}, that the doubling of $d_{eh}$ of the double SF compared to the single SF is due to the existence of the double step potentials and the small separation between two SFs. Such a shape of potential leads to approximately two-fold increase of the electric field experienced by an electron. On the other hand, the experimentally observed spread of emission energy is mainly caused by the variation of the hole confinement energy with the distance between two SFs. Comparison between the theoretical dependence, shown in Fig.~\ref{fig:hole2SF}, and the experimental distribution of the exciton emission energy suggests that the SF separation is around 4$-$6 nanometers. This conclusion is also confirmed by the STEM data on trapezoid SFs shown in Fig.~\ref{fig:haadf}(c), where the distance of $\approx 5.5$~nm between the SFs is measured.

As shown in Fig.~\ref{fig:hole2SF}, with the increase of the SF separation the hole confinement energy decreases, and thus, the exciton emission energy, see Eq.~\eqref{hw}, increases. The spread of the hole wave function also increases with larger SF separation, leading to larger electron-hole wave function overlap, and thus, the increase of the PL intensity. These conclusions qualitatively agree with the continuously distributed data ($E_{SF}>1.4891$~nm) shown in Fig.~\ref{fig:trapezoid}(a-c). A slight increase of $d_{eh}$ with the PL energy, observed in Fig.~\ref{fig:trapezoid} (c), may be attributed to increased SF separation. The origin is still not clear for the 4 discretely distributed PL energies, i.e. $E_{SF} = $1.4875, 1.4882, 1.4887 and 1.4891~eV. A plausible theory is that only certain SF separations are energetically allowed for forming stable double SF structures if the SF separation is small. This theory could be confirmed by further a correlated optical-structural imaging study of several trapezoid structures. 

\section{Conclusion}\label{sec:concl}

In conclusion, we have developed a microscopic model of the stacking-fault potential and exciton wavefunction in GaAs. Specifically, the SF potential provides a delta-function like confinement for the hole and a step-like potential for the electron. Variational method calculations for the exciton diamagnetic and magneto-Stark shifts show good agreement between theory and experiment for the single stacking-fault potential. This comparison together with DFT calculations of electronic spectrum allowed us to estimate the band offset at the SF plane as $\sim 10$~meV , which corresponds to the built-in electric field $F \sim 0.1$~MV/cm. The model also qualitatively describes the two-fold increase in the exciton dipole moment observed in the double stacking fault structure, suggesting an average inter-fault distance of 4$-$6~nm. This value is also confirmed by the STEM measurements of the trapezoid SFs.  The properties of stacking-fault excitons not only have implications for improving GaAs technologies such as solar cells and LEDs \cite{ref:guha1993dtf,ref:colli2003cns,ref:caroff2011cpt,ref:karin2016ovr}, but also provide insight into understanding potential exciton-exciton interactions and whether new excitonic phases are accessible in this system or similar systems \cite{ref:kosterlitz1973omp,ref:high2009tie}.

\acknowledgements

We acknowledge Cameron Johnson for his assistance in the experimental measurements. M.V.D. is grateful to Dr. M.O. Nestoklon for fruitful discussions and help with DFT calculations. TEM experiments and modeling were supported by Pacific Northwest National Laboratory (PNNL) Directed Research and Development program. PNNL is operated by Battelle for the Department of Energy under contract \# DE-AC05-76RLO1830. STEM imaging was performed in the Radiological Microscopy Suite (RMS), located in the Radiochemical Processing Laboratory (RPL) at PNNL. M.V.D. acknowledges financial support from the ``Basis'' Foundation for the Advancement of Theoretical Physics and Mathematics and the Russian Federation President Grant No. MK-2943.2019.2. M.M.G. and M.V.D. have been also partially supported by the RFBR grant No. 17-02-00383 and by the Program No. 13 of Presidium of RAS. K.M.C.F., X.L, and M.L.K.V. acknowledge support by the UW Molecular Engineering and Materials Center  with funding from the NSF MRSEC program (DMR-1719797). A.D.W. and A.L. acknowledge gratefully support of DFG-TRR160,  BMBF-Q.Link.X 16KIS0867, and the DFH/UFA  CDFA-05-06. 

\appendix

\section{Structural imaging}\label{sec:AppA}

Cross-sectional STEM samples were prepared using a FEI Helios NanoLab DualBeam Focused Ion Beam (FIB) microscope and a standard lift out procedure along the GaAs [110] zone-axis, with initial cuts made at 30 kV and final polishing at 2 kV. High-angle annular dark field (HAADF) images were collected on a probe-corrected JEOL GrandARM-300F microscope operating at 300 kV, with a convergence semiangle of 29.7 mrad, and a collection angle of 72--495 mrad. To minimize scan artifacts and improve signal-to-noise, drift-corrected images were prepared using the SmartAlign plugin~\cite{ref:Jones2015san} for this, a series of ten frames at $1024 \times 1024$ pixels with a 2 \textmu s px$^{-1}$ dwell time and 90$^{\circ}$ rotation between frames was used. The frames were up-sampled $2\times$ prior to non-rigid alignment, followed by template matching parallel to the fault direction. Full multislice image simulations were conducted with the PRISM code~\cite{ref:Pryor2017smi} for several candidate structures from \textit{ab initio} calculations. Simulations were performed using a $1 \times 4$ tiling for crystal thicknesses of 50, 100, and 150 u.c., corresponding to 20, 40, and 60 nm, respectively. Imaging parameters were matched to the experiment and a 0.05 \AA~px$^{-1}$ sampling, 2~\AA~slice thickness, and 10 frozen phonon passes were used for the final simulations. From these simulations, the 60 nm simulation was compared to the experiment.

\begin{figure*}[htb]
  \centering
  \includegraphics[draft=false,width=6in]{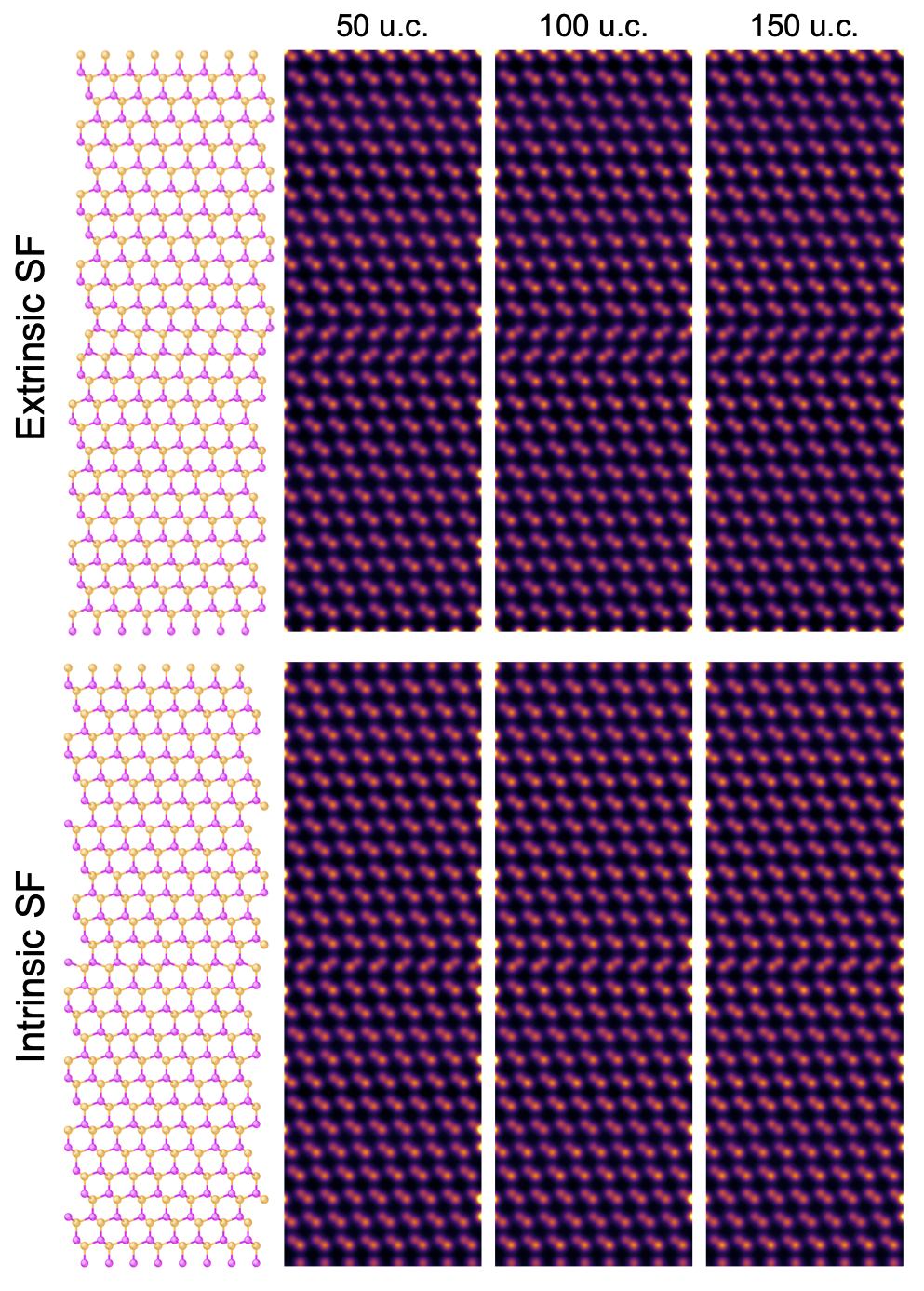}
  \caption{\label{fig:si_multislice} Series of multislice image simulations performed for the extrinsic (top) and intrinsic (bottom) stacking fault structures for 50, 100, and 150 u.c. thick crystals using the PRISM code.
  }
\end{figure*}

We have performed a series of multislice image simulations based upon our {\it ab initio} calculations for both the extrinsic and intrinsic stacking fault structures. Simulations were conducted across range of reasonable sample thicknesses, using the same experimental imaging conditions, as shown in Fig.~\ref{fig:si_multislice}. We find a good agreement between the real and simulated structures, supporting the validity of our calculations. We observe only subtle changes in image contrast with increasing thickness and find that the 150 u.c. model is most consistent with our prior knowledge of the sample and the measured data.

\section{DFT calculations of the stacking fault electronic structure}
\label{appendix:DFT}

\begin{figure*}[htb]
  \centering
  \includegraphics[draft=false,width=6in]{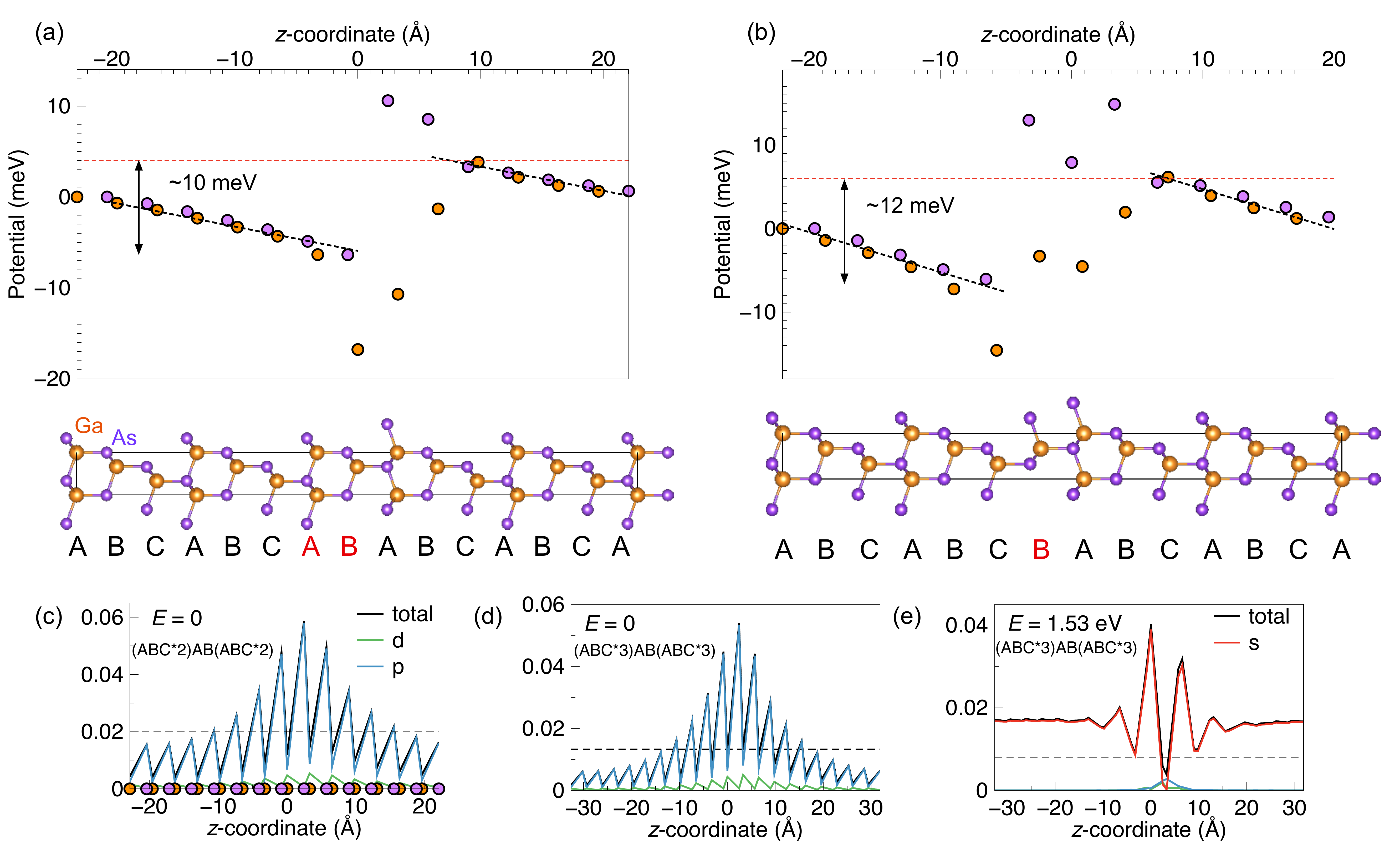}
  \caption{\label{fig:dft} Summary of the DFT calculations. (a, b) Electrostatic potential of the \textit{intrinsic} (a) and \textit{extrinsic} (b) SFs extracted from the DFT calculations. The insets in the bottom show the elementary cells used in the calculation. (c-e) Electron density of $s$, $p$ and $d$ atomic orbitals in the states with different energies in the \textit{intrinsic} stacking fault. $E = 0$ is the position of the Fermi level, and $E = 1.53$~eV is the lowest unoccupied state. The black curve is the sum of $s$, $p$ and $d$ contributions, the dashed horizontal line depicts the electron density in the interstitial (inter-atomic) region.  Panels (d, e) correspond to a stacking fault structure with six additional monolayers as compared to other panels.
  }
\end{figure*}

In order to estimate the value of the band offset $V_0$ and analyze single-electron states in the presence of the stacking fault, we performed the DFT calculations using the WIEN2k package with mBJ exchange-correlation potential~\cite{wien2k, ref:tran2009abg}. We performed calculations for two types of stacking faults, an intrinsic and an extrinsic one, that have different order of layers in the vicinity of the stacking fault, see the insets of Figs.~\ref{fig:dft}a, b. To estimate the electrostatic potential in the stacking fault structure we applied the procedure described in Refs.~\cite{ref:wei1987rov,ref:li2009rai,ref:smirnov2018ebi}, which involves tracking the position of the core 1s level at Ga and As atoms in the structure. The presented calculations were performed for relaxed structures, however we found that the value of $V_0$ is only slightly different in relaxed and non-relaxed structures. The calculations predict that the energy gap of the bulk zinc-blende phase is about 200~meV larger than that of the bulk wurtzite phase. 

The extracted electrostatic potential is shown in Figs.~\ref{fig:dft}a, b. In agreement with previous results on the ZnSe stacking faults~\cite{ref:smirnov2018ebi}, we observe an overall jump of electrostatic potential when crossing the stacking fault region. The linear behavior of the potential, i.e. non-zero electric field, outside the stacking fault region is an artifact of periodic boundary conditions used in numeric calculation. We checked that this field decreases with an increase of the elementary cell length. The oscillations of the potential, and correspondingly, of the electric field in the vicinity of the stacking fault are not eliminated by the increase of the calculation accuracy and the cell length. These oscillations reflect the atomic-scale oscillations of the charge density in the stacking fault region. The electrostatic potential change across the SF, which corresponds to the $V_0$ parameter in Eq.~\eqref{pot}, is $V_0 \approx 10$~meV and has the same sign for both intrinsic and extrinsic stacking faults. The corresponding electric field inside the SFs is $\approx 0.1$~MV/cm, which is about 5 times smaller than in ZnSe SFs~\cite{ref:smirnov2018ebi} and about 25 times smaller than in GaN SFs~\cite{PhysRevB.86.081302}.

Figures~\ref{fig:dft}(c-e) show the behavior of the electron density across the stacking fault. It is seen that the lowest state in the conduction band is delocalized, whereas the highest state in the valence band is localized with the localization length $a_h \approx 40$~\AA~(full width of density at $1/e^2$). We checked that $a_h$ does not depend on the supercell size, see Figs.~\ref{fig:dft}d,e. Hence, the stacking fault tightly binds a heavy hole and does not localize an electron, in agreement with the suggested model potential~\eqref{pot}. Using $m_{h,\perp} \approx 0.95 m_0$~\cite{ref:ikonic1992vss}, the hole confinement energy $\varepsilon_h = 2 \hbar^2/(m_{h,\perp} a_h^2) \approx 10$~meV. The energy shift between the bulk exciton and stacking-fault exciton is $\varepsilon_{h}+V_0 \approx 20$~meV, which agrees well with the experimental value $\approx$ 19 to 22~meV.

%\bibliography{ioffe.bib,xiayu.bib,maria.bib,xiayu2.bib,Maria_Bib.bib}

%bibliography section

%Control: production of eprint (-1) disabled
%

\end{document}